# Multipacting studies of $\beta_g$=0.61 and $\beta_g$=0.9, 650 MHz Elliptical SRF Cavities

Ram Prakash, Arup Ratan Jana and Vinit Kumar

Accelerator and Beam Physics Laboratory, Raja Ramanna Centre for Advanced Technology, Indore

*Abstract*

Multipacting is a resonant process, where a large number of unwanted electrons grow rapidly at some specific locations in an rf cavity. It is a parasitic discharge, which degrades the performance indicators (*e.g.* the quality factor ($Q$) and the maximum achievable accelerating gradient($E_{acc}$) ) and in case of superconducting radiofrequency (SRF) cavity, it leads to the quenching of superconducting material. Therefore, multipacting is seen as a performance limiting phenomenon for the SRF cavities. Numerical simulations are essential to pre-empt the experimentally observed multipacting conditions. Readily available codes(*e.g.* `FISHPACT, MULTIPACT, CST` etc.) are widely used to simulate the phenomenon of multipacting in such cases. Most of the contemporary two dimensional (2D) codes are unable to detect multipacting in the elliptical cavities because they use a simplistic secondary emission model, where a constant value of emission kinetic energy of secondary electrons is assumed.  Some three-dimensional (3D) codes, which use a more realistic and sophisticated secondary emission model (Furman model) by following a probability distribution for the emission kinetic energy of secondary electrons, are able to correctly predict the occurance of  multipacting. These 3D softwares however require large data handling and are slower than contemporary 2D codes. We have developed a 2D code for the multipacting analysis for axisymmetric structures, which is based on the Furman model. Since our code is 2D, it is faster than the 3D codes. This makes it possible to perform an accurate multipacting analysis of multipacting within a resaonalble computational time, thus achieving an accuracy as good as in the case of 3D codes, but requiring significantly less computing time.

We have used our code for determining the multipacting prone range of values for $E_{acc}$ in $\beta_g = 0.61$ and $\beta_g = 0.9$, 650 MHz mid-cell elliptical cavities that we have recently designed for the medium and high energy section of the proposed Indian Spallation Neutron Source (ISNS) project.

## I. Introduction

Superconducting radiofrequency (SRF) cavities are used for accelerating charged particles in high energy and high power accelerators. SRF cavities provide very high accelerating gradient ($E_{acc}$) in continuous wave (CW) mode with smaller heat loss and are therefore preferred over normal conducting cavities for high average power operation. One of the performance limiting phenomenon in such cavities is the phenomenon of multipacting, which is a resonant phenomenon that results into the building up of a large current within a small region of the cavity. Experimental observations also confirm that the onset of multipacting restricts the raising of accelerating gradient in an SRF cavity beyond a certain limit [1,2]. As the multipacting starts, the excess heat is deposited at certain location on the cavity surface, which results in quenching of the superconducting cavity. As a result of quenching, the quality factor of the cavity decreases very sharply. Multipacting is an unwanted complex phenomenon and needs to be avoided during the operation of a cavity. In this paper, we have performed a detailed analysis to develop an insight of this complex process and calculated the threshold for onset of multipacting in TESLA [3] type niobium based superconducting cavities. An indigenous computer code has been developed, which is based on a general architecture and can be modified to study the multipacting in other RF structures with complex geometry, such as- waveguides, power couplers etc..

Multipacting is generally triggered by a few seed electrons, which are already present inside the cavity due to cosmic rays, field emission, secondary emission, or beam itself. The seed electrons get energized with the cavity RF field and collide with the cavity wall. If the colliding electrons have a kinetic energy within a specific range, secondary electrons may get ejected from the cavity surface during the collision. If the motion of the secondary electrons is synchronized with the RF field and the yield of secondary emission ($\delta$) is >1, there is a build up of multipacting electrons. Here, $\delta$ is defined as the ratio between number of secondary electrons ($N_s$) ejected and the number of primary electrons ($N_p$) incident (*i.e.* $\delta = N_s/N_p$). In case of multipacting, the number of secondary electrons increases exponentially to very high value within a few RF cycles. These electrons absorb the RF energy and deposit this energy in the form of heat on the cavity surface. Experimental observations show that the heat deposition raises the temperature at the outer surface by ~100 mK, and this leads to the quenching of superconductivity of cavity surface [1]. For example, in CEBAF shape elliptical 1.5 GHz SRF cavity, multipacting was experimentally observed at peak electric field ($E_{pk}$) ~ 30-34 MV/m [1]. Historically, to reduce the problem of multipacting, shape of SRF cavities has evolved in the last few decades from pill-box to elliptical [4]. Although the modification in the shape diminishes multipacting to a large extent, the problem still persists near the equator region of a well optimized elliptical cavity [5, 6]. In our work, we have investigated the multipacting phenomenon in TESLA type elliptical SRF cavities and have shown that multipacting can be suppressed if the shape of elliptical cavity is slightly modified near the equator region [7].

Due to the complex structure of TESLA type cavities, analytical methods become difficult to implement. Therefore, numerical codes are used for multipacting analysis in such cases. A brief review of such computer simulation codes is given in Ref. [8]. Among these available simulation packages, `MultiPac` [9] and Computer Simulation Technology- Particle in Cell (`CST-PIC`) [10] are two major computer codes which are used for multipacting analysis. `CST-PIC` is a three dimensional (3D) software, which implements Furman model [11] to simulate the secondary emission process that generates a distribution of kinetic energies of secondary electrons after every collision. On the other hand, `MultiPac` is a 2D code with rather simplistic secondary emission model, where the emission energy of secondary electron is assumed to be a constant (~ few eV). Being a 2D software, `MultiPac` needs to handle much less data compared to `CST-PIC`, which makes it faster than the 3D codes. Experimentally, multipacting was reported to be observed in the equatorial region of TESLA 1.3GHz cavities at accelerating gradient ($E_{acc}$) in the range ~ 17-22 MV/m [5]. This experimental observation could not be predicted by `MultiPac` simulations since the assumption of simplified secondary emission model in the code is not consistent with the detailed secondary emission data obtained experimentally[12].

We have developed a computer code to study the multipacting phenomenon in an azimuthally symmetric elliptical SRF cavity. Our code tracks multiple particles in the *2D* plane; therefore it is fast in comparison to the existing 3D codes. In order to model the secondary emission process, Furman secondary emission model is used in our code. Due to these features, our code correctly predicts the possibility of multipacting in the TESLA type cavity and calculates the range of accelerating gradient for which multipacting is likely to occur. We have benchmarked our computer code by studying the multipacting in 1.3 GHz TESLA cavity and comparing the results obtained using other code and experimantal result [7]. We have then used our code to predict the onset of multipacting in the $\beta_g$ =0.61 and $\beta_g$ =0.9, 650 MHz elliptical cavity that has been recently designed for the proposed Indian Spallation Neutron Source (ISNS) project.

We would like to mention that secondary emission from any surface is composed of three different processes, namely- true secondary emission, rediffused emission and elastic emission [11], which are simulated numerically in the Furman model. Based on the experimental data [10] available in the literature, we make an important observation that the secondary emission from the niobium surface occurs predominantly via true secondary emission. Hence, we simplified the Furman model by ignoring the effect of other two processes of secondary emission namely- the rediffused emission and the elastic emission. We found that this remarkable simplification for the specific case of niobium leads to a significant reduction in the computation time, without affecting the results of the calculation.

The paper is organized as follows. In section II, the general architecture of the developed code is discussed. This section gives a detailed description of the particle tracking scheme along with the field interpolation scheme at the particle location. Section III describes an algorithm to detect the multipacting infected locations in the cavity, and to quickly estimate the range of values of $E_{acc}$ for which multipacting may occur. This is followed by a brief discussion of the secondary emission model in Section IV. In section V, we have benchmarked our computer code by analyzing the multipacting in TESLA 1.3 GHz elliptical cavity and comparing the results with Ref. [7]. Further, we used our code to study the multipacting and its suppression in $\beta_g$ =0.61 and $\beta_g$ =0.9, 650 MHz elliptical cavities, which is also presented in Section V. Finally, we have concluded the paper with some discussions in section VI.

## II. Methodology for Particle Tracking

The computer code that we have developed to study the multipacting phenomenon contains two parts. The first part performs the particle tracking calculations under the influence of cavity electromagnetic fields, and it also takes care of particle collision with the cavity wall. The second part uses a suitable secondary emission model to generate the complete description of secondary electrons such as - emission energy, emission angle *etc.* after the particle-wall collision. We will discuss the secondary emission model in detail in Section IV. Present section gives a detailed description of particle tracking algorithm used in the code. The code was developed in python [13] programming language, which is an interpreter based high-level language.

Electromagnetic field in any elliptical cavity can be most conviniently described in cylindrical coordinates ($r, \theta, z$). The operating mode for accelerating the charged particles in an elliptical cavity is the $TM_{010}$ mode, where the electromagnetic field inside the cavity has $E_r, E_z$ and $B_\theta$ components only. The particles will therefore experience the force only in the radial direction and *z*- direction. We now discuss an important point here. The electrons have relatively smaller energy and hence smaller velocity when they are emitted at the cavity surface. The radial and *z* component of the velocity *i.e.* $v_r$ and $v_z$ increases to higher values due to the forces acting in the corresponding directions. As will be discussed later, the electrons will be emitted perpendicular to the cavity surface, as well as making some angle $\alpha$ with this perpendicular, with the probability of emission proportional to $\cos(\alpha)$. The initial value of the velocity in the $\theta$ direction *i.e.* $v_\theta$ will be zero for the particles emitted perpendicular to the cavity surface and will be very small for the other particles which will be smaller in number. Since there will not be any force in the azimuthal dirction, the azimuthal coordinate will evolve according to the equation $r(d^2\theta/d^2t)=-2(dr/dt)(d\theta/dt)$. As one can infer from this equation, $v_\theta$ will always remain much less comapre to $v_r$ and $v_z$. It will thus be a good approximation to ignore the motion in $\theta$ direction and track the particles only in the $r$ and $z$ directions. We have therefore performed all our calculations in 2D, assuming that the particle motion is mainly confined within a 2D plane. This simplifies our calculations significantly.

To calculate the electromagnetic field in the 2D plane, we have used a computer code SUPERFISH, which is availaible in public domain. SUPERFISH constructs a mesh with rectangular cells for the 2D structure, and then gives field values at every node of the mesh. The mesh, with field values at every node, is imported by our code. Our code then uses these electromagnetic fields to determine the charged particle movement in the rectangular mesh. The position is determined in the mesh after every small time step. Value of the electromagnetic field is estimated at the particle location, using the first order field interpolation by considering the proximity to the mesh nodes. The interpolated field thus obtained is used to calculate the position and momentum of charged particle in the next small time step *dt*, by solving the Lorentz equation of motion. Acceleration due to Lorentz force can be written in the relativistic form [14] as

$$\vec{a} = \frac{q}{m\gamma}\left[(\vec{E} + \vec{v} \times \vec{B}) - \frac{\vec{v}(\vec{v}\cdot\vec{E})}{c^2}\right]. \quad (1)$$

Here, a particle with charge *q*, mass *m* and velocity $\vec{v}$ is accelerated by the EM field with acceleration $\vec{a}$, and $\vec{E}$ and $\vec{B}$ are the electric and magnetic field respectively at the particle location, and *c* is the velocity of light and relativistic factor $\gamma = 1/\sqrt{1 - \frac{v^2}{c^2}}$.

Being a second order ordinary differential equation (ODE), equation(1) can be decomposed into two first order ODEs. In a 2D plane, these ODEs are numerically solved using Leap-Frog method [15] and the charged particle movement is tracked in the mesh. In this method, particles movement in electromagnetic field for the time step *dt* is assumed to be taking place in three successive steps [16]. In the first step, particle is accelerated in the influence of electric field only for the first half of the time period *dt* (i.e. *dt*/2). The second step corresponds to the rotation of velocity vector of particle by the magnetic field only for time interval *dt*. In the last step, particle is again accelerated exclusively by the electric field for the remaining half of the small time period *dt* (i.e. *dt*/2). The value of position and velocity is recorded at the end of every time step *dt*. In this way, particle is tracked till it strikes the boundary of the RF cavity. The boundary for the RF cavity is established from imported field itself by distinguishing the region with non-zero field values with no field region. As soon as the particle collides with the boundary, equation solver is interrupted and collision kinetic energy and angle of incidence of the particle are calculated. Using the kinetic energy and incidence angle data, Furman model (discussed in Section IV) simulates the secondary emission process according to surface properties. In this way, we obtain a data set of kinetic energies and emission angles of secondary electrons, which are emitted from the surface. This data set is used to re-intialize the equation solver. In the next section, we will discuss how to determine the probable multipacting locations in the SRF cavity. We will also find out the range of $E_{acc}$ for which resonant trajectories are possible.

## III. Identification of the Multipacting Prone Locations and Resonant Trajectories

As discussed in Section I, multipacting occurs when the trajectory of the multipacting electrons is synchronized with the RF field such that their energy increases to values for which the secondary emission yield >1. In order to perform a detailed multipacting analysis, one needs to simulate a large number of possible scenarios. For example, the initial position of seed electron needs to be taken arbitrarily inside the cavity volume to exhaust all the possibilities. Also, the acceleration gradient in the cavity needs to be varied over the complete range. In this section, we will perform some simple calculations to find out the multipacting prone locations on the cavity surface and also find out a range of $E_{acc}$ for which multipacting is likely to occur, such that we focus our computational effort to perform detailed multipacting calculations near that location for the specified range of $E_{acc}$.

We now briefly describe the simplistic approach taken for performing such calculations. We assume the initial position of the seed electron randomly anywhere inside the cavity volume and take its initial kinetic energy to be arbitrary. The trajectory of the seed electron is traced till it strikes the cavity surface. When the seed electron strikes the surface, we assume that a secondary electron is emitted perpendicular to the surface with kinetic energy of 3.5 eV (corresponding to particle velocity of $1.11 \times 10^6$ m/s). This assumption originates from the fact that the most probable value of kinetic energy of the true secondary emission from niobium surface is found to be ~3.5 eV [10]. Note that although the emission energy of secondary electrons follow a distribution in reality, we have here taken a fixed value of emission energy in our simplistic approach. Trajectory of the emitted particle is further calculated till it strikes the cavity surface, followed by the emission of secondary electrons as described earlier, and so on.

Figure 1 shows the trajectory of electrons for two different initial locations and two different initial phases (60° and 120°) of the seed electron assuming an accelerating gradient of 24 MV/m. The trajectories are calculated for the TESLA 1.3 GHz elliptical cavity. We notice that after around 20 collisions, the trajectories settle in the equatorial region. More importantly, this is independent of initial phase and location of seed electron. We further notice that after the trajectory settles, the electrons move back and forth between two points on the cavity surface and follow a repetitive path, which is shown as a satelite image in Figure 1. Figure 2 shows a plot of phase difference between two consecutive collisions as a function of impact number. We observe that after few collisions the consecutive phase difference settles at $\pi$, which indicates that there is a possibility of two point first order multipacting [5] in this situation.

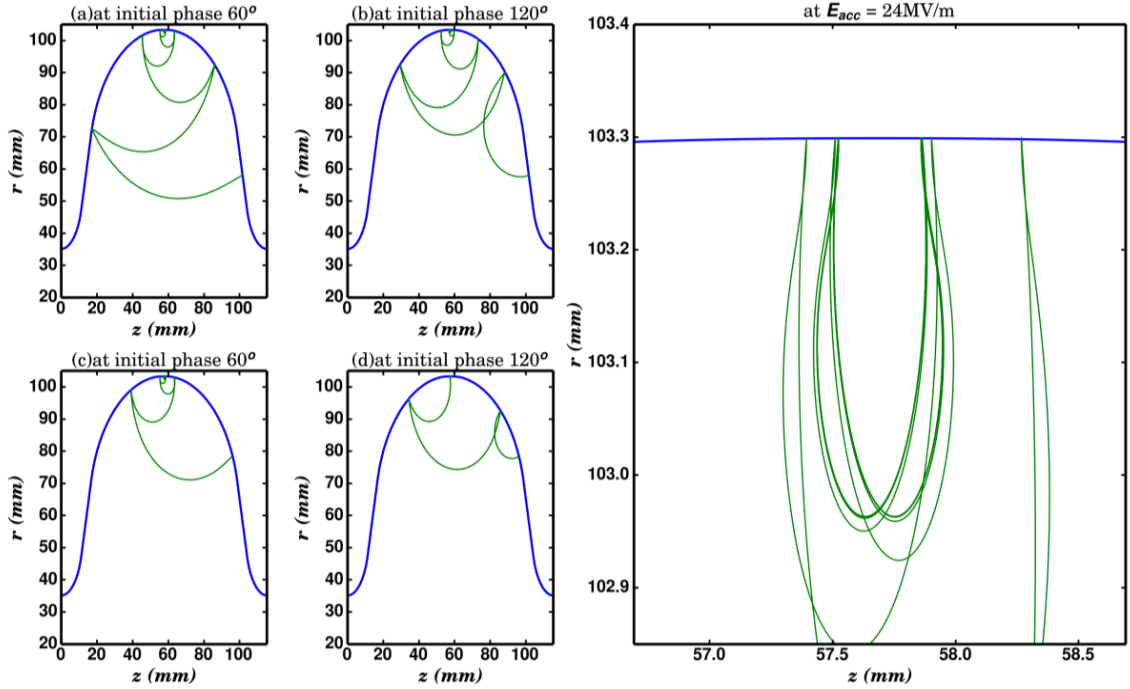

Figure 1: Trajectory of the particle at $E_{acc}$= 24 MV/m, when initial particle is emitted from two different locations and at two different phases. Initial phase has been taken as $60^o$ for the trajectories shown in (a) and (c). Similarly, the initial phase is $120^o$ for (b) and (d) . Final trajectory of the particle is shown in a zoomed view of equator region in the satelite image.

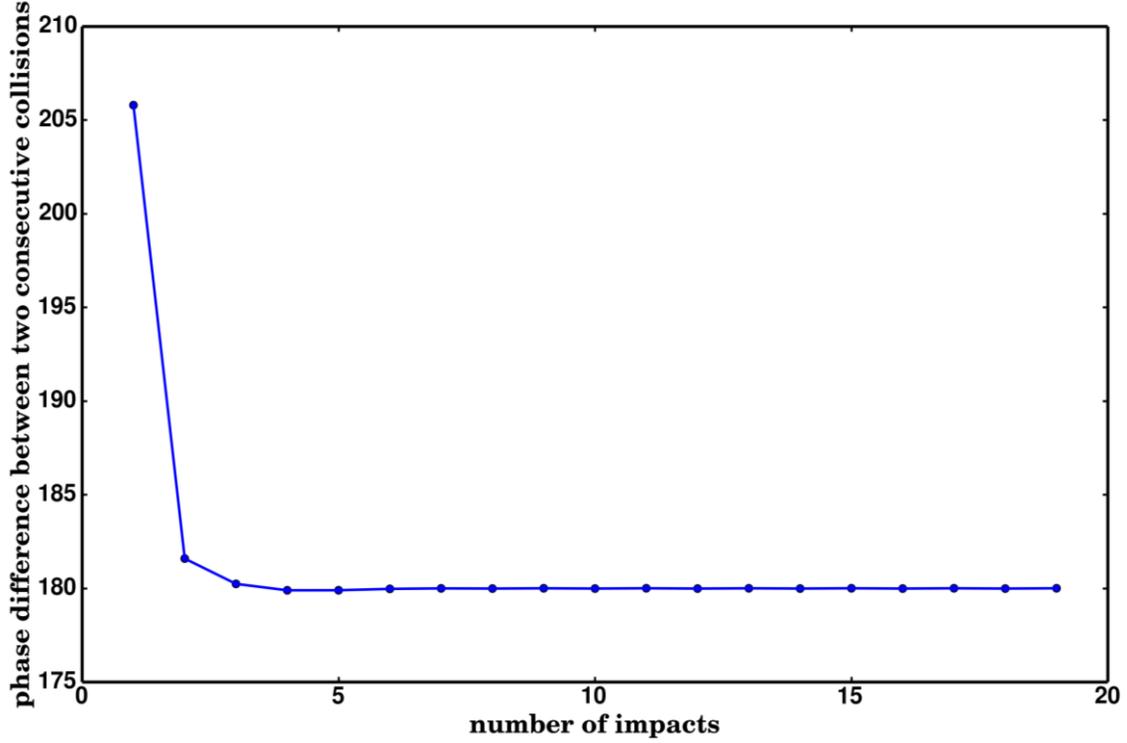

Figure 2: Phase difference between consecutive impacts as a function of impact number at $E_{acc}$=24 MV/m. The phase difference settles on ~ 180° after a few collisions.

We have performed similar calculations for different values of $E_{acc}$. Figure 3 shows the trajectory of electron for $E_{acc}$ = 13 MV/m. We once again notice that the particles, emitted from two different locations on the surface, and with two different initial phases ultimately settles near the equator region after a few collisions with the cavity wall. As the particle settles at the equator, it follows the repetitive path and strikes the cavity surface near the point of emission as shown in the satelite image in Figure 3. The time of flight for such particle is equal to the time period of the RF field. Therefore, phase difference between consecutive collisions with the wall as a function of impact number, which is plotted in Figure 4, shows that it settles to a constant value of $2\pi$ after a few impacts. This shows the possibility of one point multipacting. However, as will be discussed in the later part of this section, gain in the kinetic energy is not sufficient for this case to produce secondary electrons from the surface. Therefore, the multipacting does not build up even though the resonant trajectories are possible.

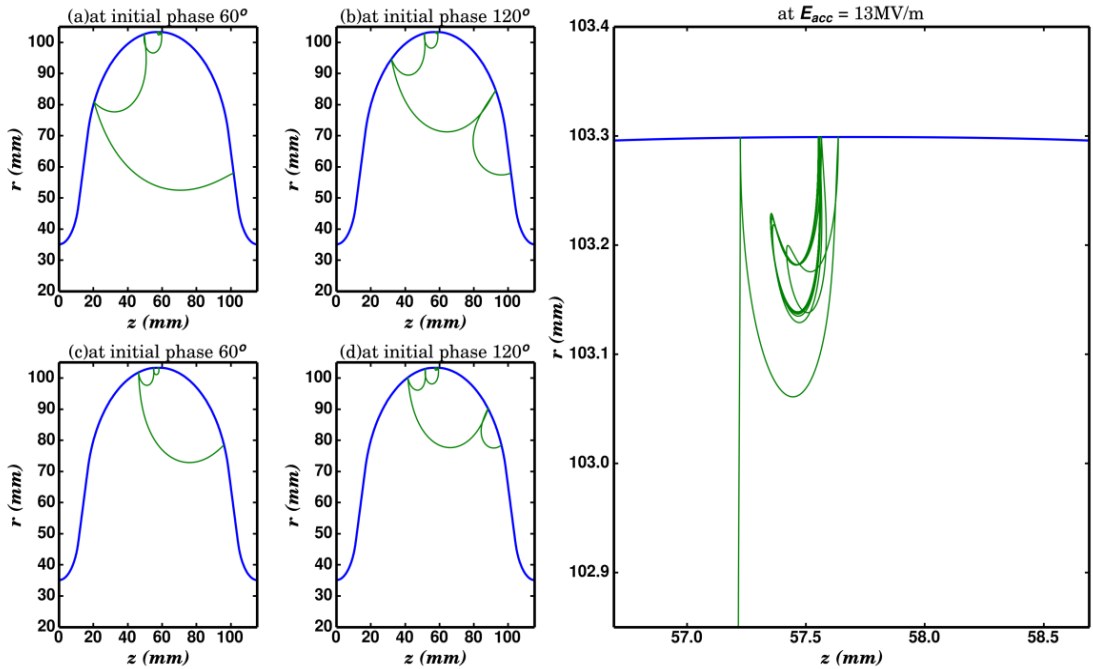

Figure 3: Trajectory of the particle at $E_{acc}$ = 13 MV/m when initial particle is emitted from two different locations and two different phases. Initial phase has been taken as $60°$ for the plots shown in (a) and (c). Similarly, initial phase is $120°$ for (b) and (d). Final trajectory of the particle is shown in a zoomed view of equator region in the satelite image.

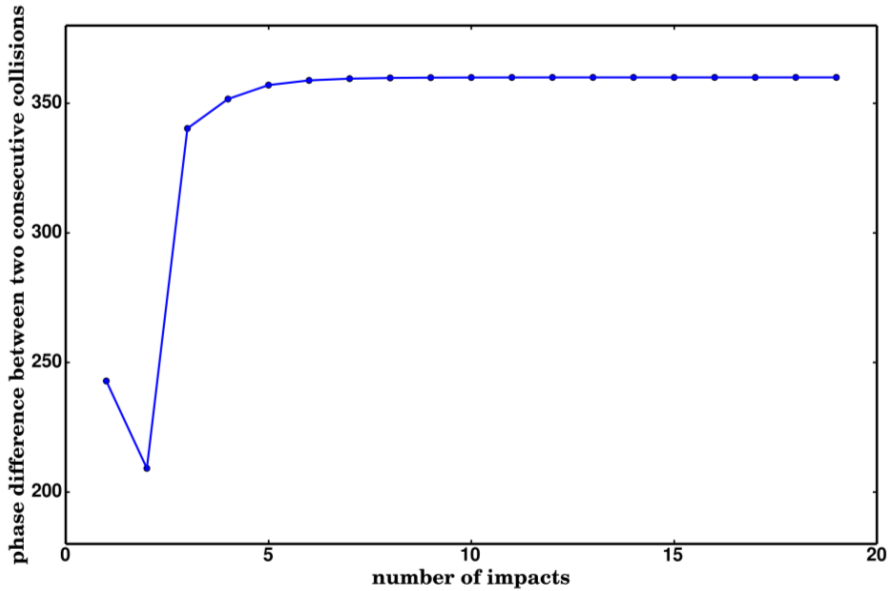

Figure 4: Phase difference between consecutive impacts of particle as a function of impact number with cavity wall at $E_{acc}$=13 MV/m.

In a similar way, we have analyzed trajectories at $E_{acc}$ = 35 MV/m. In Figure 5, we can see that the particles, which are emitted at various locations and different phases still settle at equator after few collisions, except for one case, where it goes towards equator and comes out through the beam pipe. Thus, although there is a tendency for the trajectory to settle near the equator region, there are some instances where the the electrons starting from locations near the iris travels towards the beam axis and most likely get lost in the beam pipe. These electrons do not participate in multipacting near the equator region. Attraction towards the equator region or towards the beam axis, can be explained using the pandermotive forces [17]. In Figure 5, the satellite image shows an enhanced view of the equator region and trajectory of the particle. We observe that the trajectory is complicated and cannot be defined either as one point multipacting or two point multipacting. In Figure 6, The phase difference between two consecutive collisions is plotted. We can observeve that the particle repeats the trajectory after three collisions. This type of trajectories was seen and reported for paralell plate structres and are termed as ping-pong modes[18].

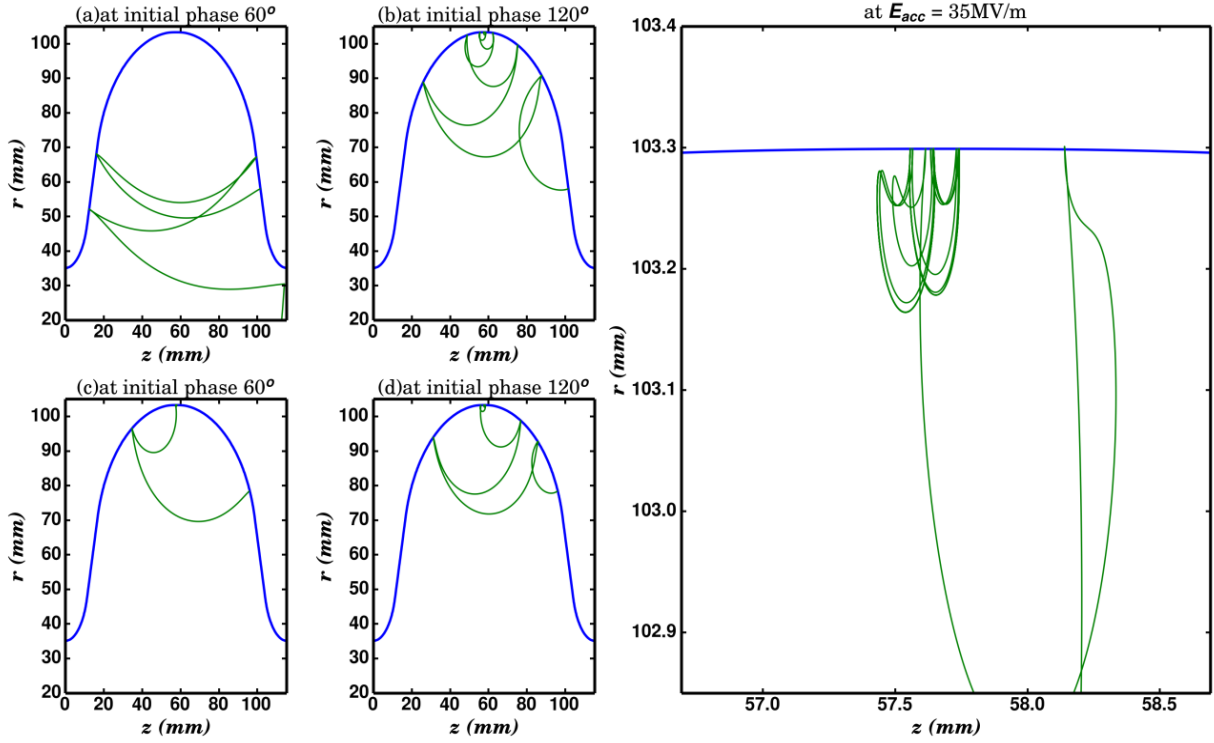

Figure 5: Trajectory of the particle at $E_{acc}$= 35 MV/m, when initial particle is emitted from two different locations and at two different phases. Here, trajectories shown in (a) and (c) are plotted when initial phase is taken as $60^o$. Similarly, (b) and (d) are plotted when initial phase is taken as $120^o$. A zoomed view of the trajectory near the equator region is shown in the satelite image.

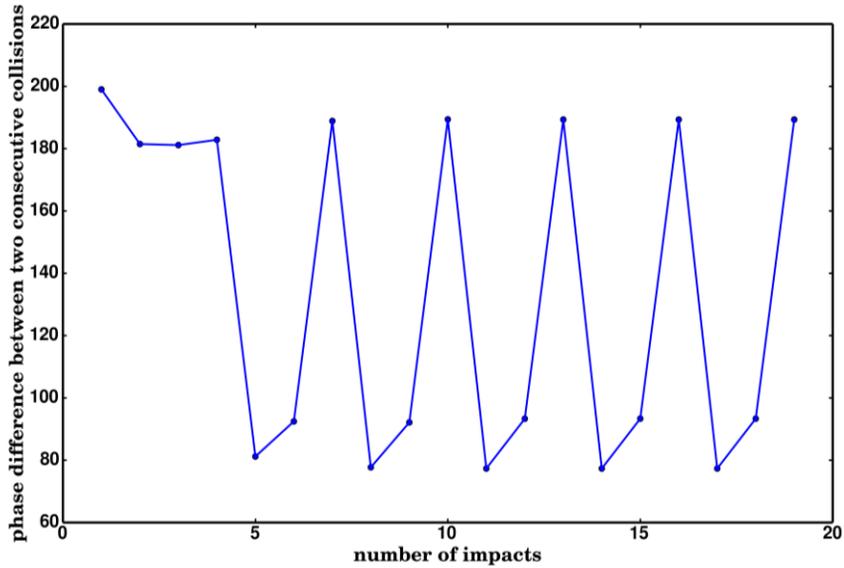

Figure 6: Phase difference between consecutive collisions is plotted as a function of number of impacts with the cavity wall at $E_{acc}$=35 MV/m.

In Figure 7, we have plotted the gain $g$ in kinetic energy of the electron when resonant conditions are satisfied. Here, $g$ can be defined as a ratio of kinetic energy just before collision and the initial kinetic energy. We can see that there is huge gain ($g>2$) in the kinetic energy of electrons in the range ∼15-31 MV/m. Interestingly, this range coincides with the range of accelerating gradient for which resonant trajectories corresponding to two point first order multipacting exist. This gives a possible indication that in this range of accelerating gradient, sustained multipacting may occur. We will explore this in the next part of our calculation. Gain in the lower gradient values $E_{acc}$ < 15 MV/m is very small, which implies the absence of multipacting for the case of one point multipacting even though the particle experiences resonanace with the RF field. At higher gradients $E_{acc}$ > 31MV/m, we observe that electrons follow a complicated resonance pattern and gain sufficient kinetic energy. Detailed investigations therefore need to be done to further explore the possibility of multipacting for this case. However, SRF cavity is more likely to attain thermal runaway and finally may get quenched before reaching such acceleration gradients.

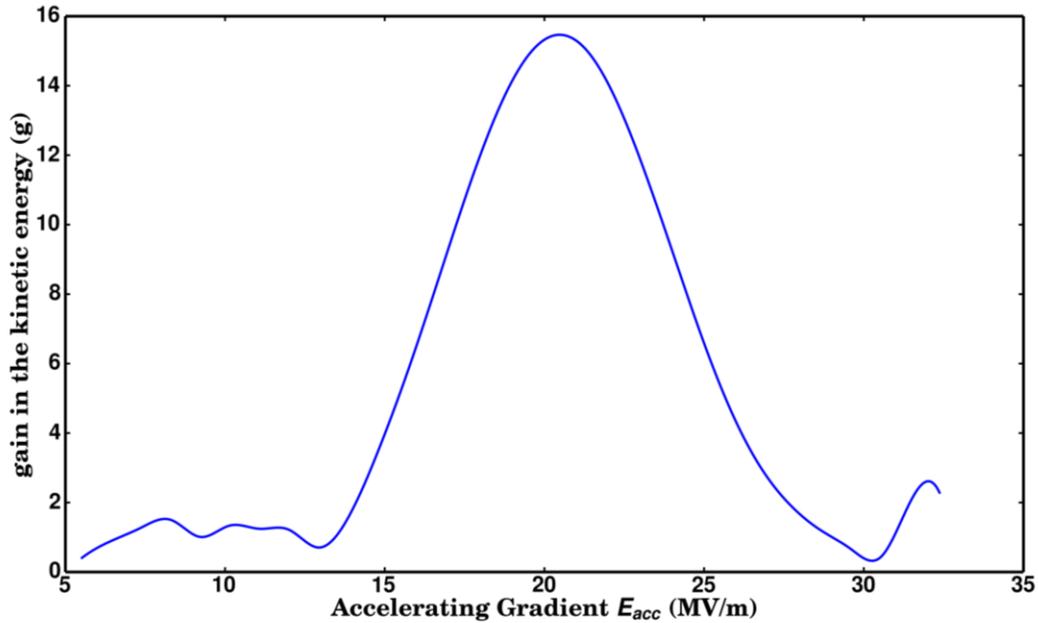

Figure 7: Gain in the kinetic energy of multipacting electrons when secondary electron is emitted with initial kinetic energy = 3.5 eV for TESLA 1.3 GHz elliptical cavity.

In Figure 8, we have shown a 2D colour plot, where the radial direction shows variation of $E_{acc}$ and angular direction shows the variation of initial phase, and the average phase difference between consecutive collisions is color coded on the 2D plane. For the region coded with light sky blue colour corresponding to the case 14 MV/m<$E_{acc}$<31 MV/m, the average of phase difference between consecutive surface collision is $\pi$. This indicates the presence of resonant trajectories corresponding to two point multipacting. We therefore focus our computational effort to study multipacting for this particular range of acceleration gradient in the next section, where we use the Furman model to simulate the secondary emission process when the electron collides the cavity wall at the end of its trajectory.

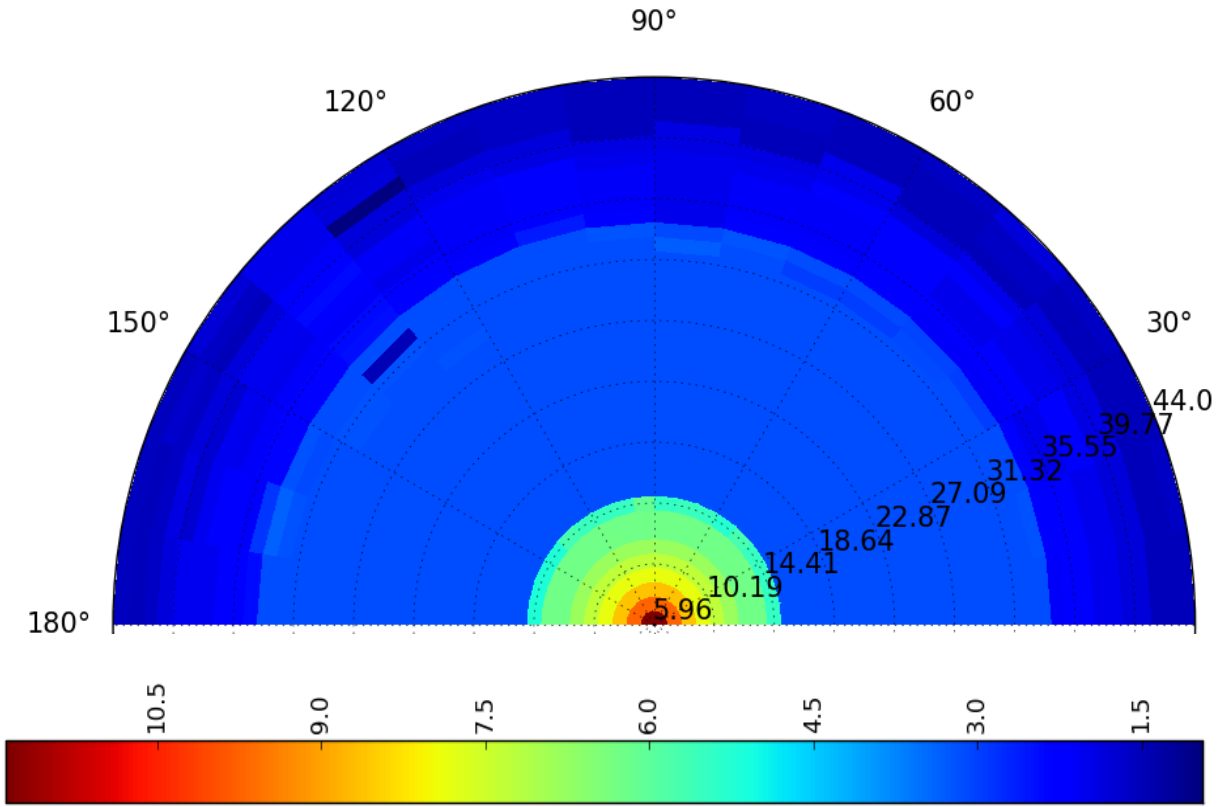

Figure 8: Colormap of average phase difference between consecutive collisions as a function of $E_{acc}$ and initial phase $\phi_0$. The region, where average phase difference is $\pi$, shows presence of resonant trajectories corresponding to first order two point multipacting. This colormap is generated for the TESLA 1.3 GHz Elliptical SRF cavity.

## IV. Secondary Emission Model

In the previous section, we used an approximate model to discuss the trajectory of the particle colliding with the cavity wall under the influence of electromagnetic field inside the cavity. This gives us an insight regarding the possibility of resonant trajectories for certain range of $E_{acc}$, which may lead to multipacting. In this section, we make our calculations more realistic by including the details of the secondary emission process to predict the growth of number of particles due to multipacting. If motion of multipacting electrons is in resonance with the RF field and overall secondary emission yield is >1, then the number of electrons will grow exponentially. In this section we have included the details of secondary emission process using Furman Model, which is implemented in our computer code.

### Furman Model

The secondary emission model is needed for a more accurate description of the multipacting process. While we track every individual electron till it collides with the surface, we require such a model to get the number of emitted secondary electrons, their kinetic energies and direction of emission. Also, it must be ensured that the emission model should produce the results, which agree with the experimentally observed data. Furman and Pivi [11] established such a secondary emission model based on the probabilistic description of the emission of secondary electrons. The Furman model was originally implemented in computer code to study the electron cloud effect (ECE) in storage rings [11]. Since this model provides a detailed description of a secondary emission process initiated by the collision of each primary electron with the surface, and is easily implementable in computer codes, we have used it in our analysis also. In this model, the contribution to the secondary emission yield is assumed to be arising from three distinct channels, with which primary electron interacts with the surface. If the incident primary electron strikes at the surface and bounces back elastically with the same energy, it is known as elastic secondary emission. If the primary electron collides inelastically with the surface atoms and is emitted with lesser kinetic energy, this is known as re-diffused emission. The incident electron can penetrate into the solid and may interact with the atoms inside in a complicated manner, resulting in emission of electrons; such an emission is known as true secondary emission. Final secondary emission data gets contribution from all the three secondary emission processes. Mostly, contribution from true secondary process is prominent. However, for smaller primary energies, contribution from the elastic and rediffused electrons is comparable to the true secondary contribution. Unlike for elastic and re-diffused case, energy distribution of emitted electrons in the true secondary emission is independent of the direction of incident electrons and mean emission energy of secondary electrons is independent of the kinetic energy of incident primary electron. Keeping the three emission processes in mind, we can write

$$\delta = \delta_{elastic} + \delta_{re\text{-}diffused} + \delta_{true\text{-}secondary}, \qquad (2)$$

where $\delta$ is the overall secondary emission yield (SEY) from the surface, which consists of contribution of elastic emission yield ($\delta_{elastic}$), rediffused secondary emission yield ($\delta_{re\text{-}diffused}$) and true secondary emission yield ($\delta_{true\text{-}secondary}$). Similarly, the secondary emission spectra $d\delta/dE_s$, where $E_s$ is the secondary emission energy can also be assumed to be comprising of the three processes. Secondary emission yield can be defined as the area under curve for the total secondary emission spectra ($d\delta/dE_s$):

$$\delta(E_0, \theta_0) = \int_0^\infty \frac{d\delta}{dE_S}(E_S, E_0, \theta_0) dE_S \qquad (3)$$

Here, $E_0$ is the kinetic energy of primary incident electron at an angle $\theta_0$ with normal to the surface. Usually, a secondary electron can get emitted by any one of the three processes. However, the probability of individual process is a function of the kinetic energy and the angle of incidence of primary electron. The purpose of Furman model is to find out the joint emission probabilities of secondary emission from all the three emission processes. These probability distributions follow basic constraints for energy conservation *e.g.* the kinetic energy of any individual secondary electron, as well as the sum of kinetic energy of all secondary electrons cannot exceed the kinetic energy of primary electron.

The probability distribution incorporates the emission probability due to all the three processes (i.e. elastic, re-diffused and true secondary) individually. Though experimental data for $\delta$ and $d\delta/dE$ is available in plenty, Furman model does not use the raw data directly. Experimental data is fitted with several parametric equations, to generate a set of suitable parameters, which describe the experimental data in the best possible manner. These parameters are used to calculate the required probability distribution. Further, Monte-Carlo simulation technique is used to obtain the computer generated secondary emission events. Here, Monte-Carlo method does not simulate the microscopic detail, but models the secondary emission event directly to obtain number of secondary electrons, along with respective kinetic energies and emission angles.

In our code, the parameters to generate the experimental data for $\delta$ and $d\delta/dE$ are given as an input to the Furman Model. When the code is executed for a large number of secondary emission events, using a given value of primary kinetic energy and angle of incidence, it is crosschecked that the data for $\delta$ and $d\delta/dE$ generated by the code agrees with the experimental data. This verifies that code can be used as an authentic secondary emission event simulator.

## Simplified model

In this subsection, we describe an approximation to the model described in previous subsection, which greatly simplifies the calculations and makes it possible to do the multipacting calculations by tracking a single particle instead of large number of particles. As we have discussed in Section I, secondary emission from the niobium (Nb) surface mainly occurs through true secondary process. We can therefore implement the secondary emission model without considering the effect of other two emission processes, namely- elastic and inelastic emission for performing simple calculations first. As a result of this, the secondary emission spectra from the Nb surface gets simplified as shown in Figure 9(a). Here, we assume that the electron is incident perpendicular to the surface. This spectra can be represented by the functional for given by $E_s^\alpha \cdot \exp(-\beta E_s)$ [11], where $E_s$ is emission kinetic energy of secondary electron, $\alpha$ and $\beta$ are parameters, which are determined from the experimental data. In our case, we have taken $\alpha = 0.498$ and $\beta = 0.133$ eV$^{-1}$, which is extracted from the data given in Ref. [11]. Here, $\alpha$ and $\beta$ decide the value of energy ($=\alpha/\beta$) at which the spectrum will be peaked. Thus, we can denote the emission spectrum as

$$n(E_s, E_0) \equiv c(E_0) \cdot E_s^\alpha \cdot \exp(-\beta E_s). \tag{4}$$

Here, $c$ is the normalization coefficient, which can be determined by requiring that

$$\int_0^\infty c E_s^\alpha \cdot \exp(-\beta E_s) \, dE_s = \delta(E_0). \tag{5}$$

This gives us

$$c \equiv \frac{\beta^{\alpha+1} \delta(E_0)}{\Gamma(\alpha+1)}. \tag{6}$$

Here, $\Gamma$ function can be defined as

$$\Gamma(t) = \int_0^\infty x^{t-1} \exp(-x) \, dx. \tag{7}$$

In a practical case, when a bunch of primary electrons strikes the material surface, another bunch of the secondary electrons is emitted with initial energy following the the statistical distribution described by Equation (4). This bunch will travel within the cavity volume under the influence of RF electromagnetic field and will strike at another place on the cavity surface. Just before striking the surface, this bunch will contain a large number of particles, which will have different kinetic energies and energy for every electron is $g$ times the energy, which it had during emission, as shown is shown in Figure 9(b). When the accelerated electron bunch strikes the cavity wall, each electron in the bunch will generate next secondary electron bunch from the surface with an initial energy distribution described by Equation (4). We thus calculate the initial energy distribution of the resulting electron bunch, which is shown in Figure 9(c).

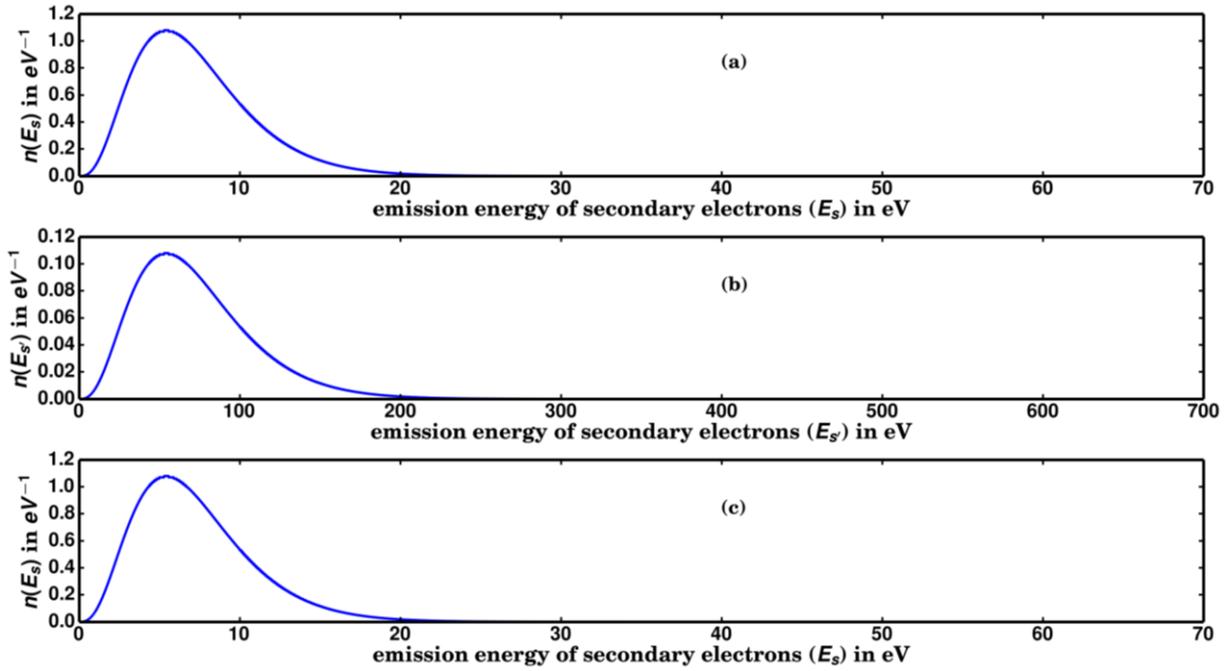

Figure 9: The emission spectrum of secondary electron bunch, just after the emission is shown in (a). The energy distribution of the electron bunch, after accelertaion due to electromagnetic field of the cavity, is shown in (b). Here we assume that gain $g$ in the energy is 10. The plot in (c) shows the secondary emission spectra of the emitted electrons when electron bunch with energy distribution shown in (b) strikes the surface.

As discussed earlier, gain in the kinetic energy of individual electron when it strikes the cavity surface at another location after getting accelerated by RFfields, is by a factor $g$. Therefore, the energy distribution of the electron bunch just before the collision can be described as $\chi \cdot \left(\frac{E_{s'}}{g}\right)^{\alpha} e^{-\frac{\beta E_{s'}}{g}}$. Here, $\chi$ is a constant. Further, every particle of the bunch will perform motion under the influence of cavity electromagnetic field and then again strike the cavity surface to generate secondary electron bunches, and so on. A final emission spectra can be calculated as a linear combination of the emission spectras from these individual primary electrons, at different incident kinetic energies. It can be written as following

$$\int_0^\infty \int_0^\infty \left[ \left( \frac{\left(\frac{\beta}{g}\right)^{\alpha+1}}{\Gamma(\alpha+1)} \right) E_{s'}^\alpha e^{-\frac{\beta E_{s'}}{g}} \right] \frac{\beta^{(\alpha+1)} \delta(E_{s'})}{\Gamma(\alpha+1)} E_s^\alpha e^{-\beta E_s} dE_{s'} dE_s = \kappa\ (g)\ , \qquad (8)$$

where $\kappa$ is the overall secondary emission yield after every collision and we observed that this independent of initial distribution of the particle and depends only on $g$. Using the relation between $g$ and $E_{acc}$ in Figure 7, we get secondary emission yield as a function of accelerating gradient ($E_{acc}$). Further, the growth rate is calculated from the secondary emission yield as $(\kappa - 1)/(T_{diff})$, where $T_{diff}$ is the time difference between consecutive collisons, which can be obtained from Figure 8. Using this method, in the next section, we have calculated the growth of multipacting electrons as a function of $E_{acc}$ in TESLA 1.3 GHz elliptical cavity, as shown in Figure 10. The calculated range of $E_{acc}$ for which multipacting occurs is observed to be ~18.5 - 26 MV/m.

## 5. Benchmarking and Results

As we have discussed in Section III, the equator region of elliptical cavity is a multipacting prone region, where the electric field has a minimum and the magnetic field is high. Taking the example case of TESLA cavity, we have seen that any seed electron, which is initialized at random location, slowly drifts towards the equator region and finally settles down there after a few RF oscillations. Therefore, at the equator region, number of electrons increases exponentially to a very high value, when the multipacting occurs. Thus, the field values inside the cavity were calculated using SUPERFISH with an special attention to eaquator region by meshing this area very dense. As resonant trajectories are very small, having an average radius of curvature around few mm, mesh spacings are chosen five times smaller for better results. We have focuused our computational effort for the range of $E_{acc}$, where resonant trajectories are found for the two point first order multipacting, as discussed in Section III. To start the simulation, we have initialized a set of 100 elctrons within a small window of 4 mm×4 mm near the equator region. These electrons are initialized with uniformly distributed random position and velocity . Here, we would like to mention that we have performed the complete simulation by increasing the number of initial electrons 10 times and final results are similar to the case when the code is initialized with a 100 electrons. Hence, to keep simulation time small, we fixed the number of initial electrons at 100. As discussed in the previous sections, the indigenous computer code first tracks the particles inside the RF cavity and then simulates secondary emission process during the particle wall collision. After every such collision, number of total mulatiapcting electrons is calculated. In this way, we can determine the number of multipacting electrons as a function of time. To present the final results for an SRF cavity in a succinct way, we have calculated the growth rate $dn/dt$ of multipacting electrons as a function of $E_{acc}$. The growth rate can be calculated by fitting the $n$ vs $t$ data numerically with an exponential of the form $n=a. \exp(bt)$, $a$ and $b$ are constants determined from the least square regression [19]. Here, $b$ is known as growth rate of the multipacting electrons. Similarly, $n$ vs $t$ data for other values of $E_{acc}$ were produced and fitted with an exponential. For a particular field gradient, a positive growth rate means that the number of multipacting electrons increases exponentially and therefore multipacting is obsereved . Similarly, negative growth rates mean that the number of particles decreases exponentially and hence multipacting will not be obsereved in such cases. We have used the developed computer code to study the multipacting in the TESLA 1.3 GHz elliptical cavity. In the Figure 10, we have plotted growth rate ($b$) is as a function of $E_{acc}$ for TESLA 1.3 GHz elliptical cavity.

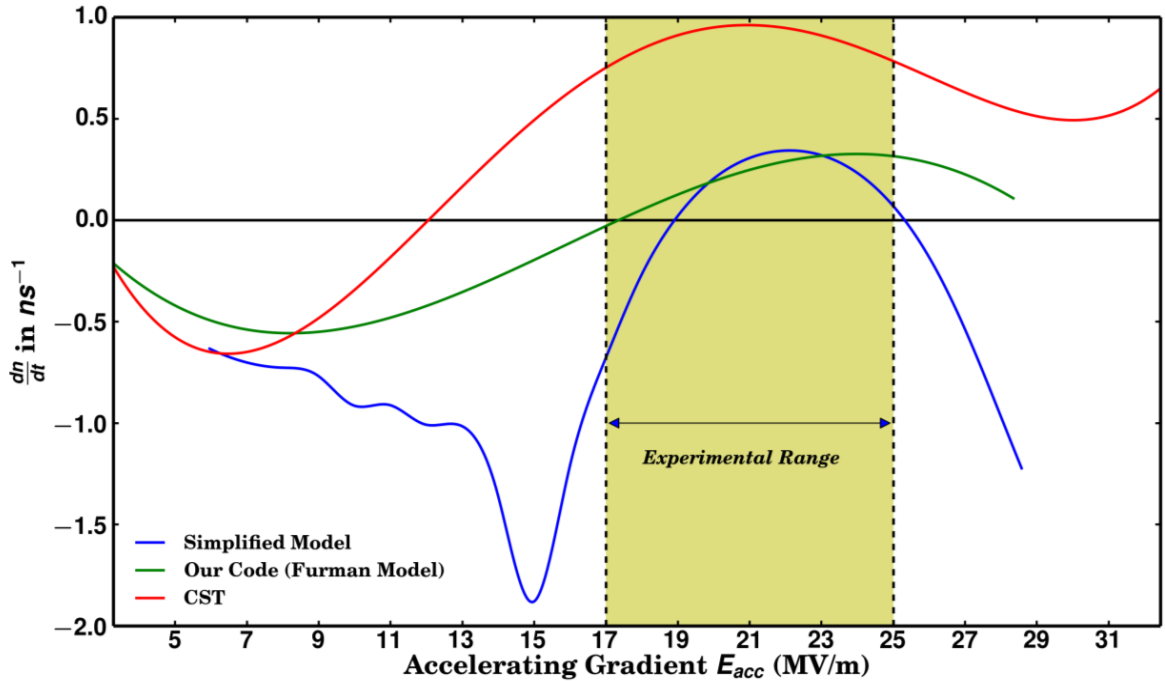

Figure 10: Growth rate (*dn/dt*) of the number of multipacting electrons is plotted as a function of $E_{acc}$ for TESLA 1.3 GHz elliptical cavity. These plots have been obtained using simplified model, using our code and CST.

We can see in Figure 10 that the onset of multipacting occurs for $E_{acc}$ ~ 17.5 MV/m. In the TESLA 1.3 GHz cavity, the onset of multipacting was experimentally observed for the range of $E_{acc}$ ~ ~17-25 MV/m [5,6,7]. We have also performed the multipacting analysis of TESLA 1.3 GHz cavity using the computer code `CST-PIC`, which is a commercially available 3D software that uses Furman Model as the secondary emission model. We have used CST-Microwave Studio to calculate the field values in $TM_{010}$ resonance mode of the cavity. Field values are then exported to the `CST-PIC`. The particle emission source was defined on the surface of densely meshed region with initially 1000 particle emission. The growth rate is calculated from *n* vs *t* data and plotted in Figure 10. The onset of multupacting is seen much earlier at $E_{acc}$ ~12 MV/m in the results obtained from CST, which has small deviation in comparison to the results obtained using our code. As discussed in the previous section, we have calculated the grwoth rate using the simplified model. The onset of multipacting in this case is seen at $E_{acc} \sim 19$ MV/m. Here, we observe that simplified model correctly calculates the range of $E_{acc}$ for which multipacting is likely to occur. This range is consistent with the range obtained using indigenous code and the experimental data. In further calculations, we have used only indigenous code and CST for multipacting analysis.

## Multipacting in $\beta_g$ =0.61 and $\beta_g$ =0.9, 650 MHz mid-cell cavities

After having benchmarked our procedure for the analysis of multipacting, we next performed the calculations for $\beta_g$ = 0.61 [20] and $\beta_g$ =0.9 [21], 650 MHz, niobium based superconducting elliptical cavities, which are designed for H⁻ particle acceleration in the medium and high beta section of the proposed Indian Spallation Neutron Source (ISNS) project [20]. Theses cavities will be used to accelerate the H⁻ particle from 160 MeV to 1 GeV. The indigenously developed code is used to determine the range of $E_{acc}$ values for which multipacting occurs in the $\beta_g$ =0.61 and $\beta_g$ =0.9 elliptical SRF cavities. Since these cavities have cells with elliptical geometry, we have primarily focussed our computational effort around the equator region, where the multipacting is most probable. We have calculated the possible resonant trajectories using the procedure described in Section III. In the $\beta_g$ =0.61, 650MHz SRF cavity, we have also plotted the phase difference between the consecutive particle collisions as a function of $E_{acc}$ for the initial phase in the range $0° < \phi_0 < 180°$, in the same way as done for Figure 8 for the case of TESLA 1.3 GHz cavity. We realize that the resonant trajectories related to the two point first order multipacting are found for $E_{acc}$ in the range 8 MV/m to 16 MV/m. Using this range of $E_{acc}$ as the initial guess, we have used our computer code for the detailed multipacting analysis. Figure 11 shows the growth rate ($dn/dt$) as a function of $E_{acc}$ for $\beta_g$ =0.61, 650 MHz cavity. We can see that $E_{acc}$ has values in the range ~ 9.2-13.3 MV/m, for which multipacting occurs. Similarly, multipacting analysis is also performed for the $\beta_g$ =0.9, 650 MHz cavity. In the Figure 13, growth rate ($dn/dt$) for number of multipacting electrons ($n$) is plotted as function of $E_{acc}$. Multipacting prone range of $E_{acc}$ values for this cavity is obtained as $E_{acc}$ ~ 9.7-14.5 MV/m. In our case, $\beta_g$ =0.61 and $\beta_g$ =0.9 cavities are designed to be operated at $E_{acc}$ ~15.4 MV/m [20] and $E_{acc}$ =18.5 MV/m, respectively [21]. Hence, both the cavities may get infected by multipacting during their operation. Therefore the multipacting needs to be suppressed in this cavity to operate at the design value of $E_{acc}$. The suppression procedure of multipacting is discussed later in this section. We have also used `CST-PIC` code for multipacting analysis of $\beta_g$ =0.61 and $\beta_g$ =0.9 cavities. Multipacting is found at $E_{acc}$ ~ 7.2-13.5 MV/m for $\beta_g$ =0.61 cavity and $E_{acc}$ ~ 7.7-17.8 MV/m for $\beta_g$ =0.9 cavity, using the `CST-PIC` code.

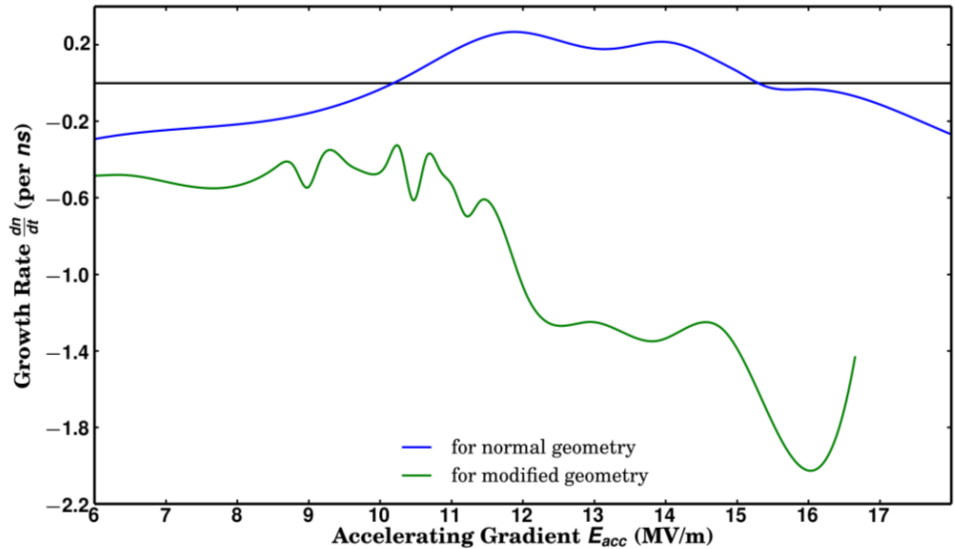

Figure 11: Growth rate ($dn/dt$) of multipacting electrons is plotted as a function of $E_{acc}$. Growth rates for the unmodified cavity is shown by the curve in blue color, while growth rates for the modified geometry is shown by the curve in green color. These results are obtained for the $\beta_g$ =0.61 elliptical cavity using the indigenously developed computer code.

## Suppression of multipacting

Secondary electrons in the elliptical cavity have tendency to settle near the equator region, where the minimum of electric field is present. Electrons at the equator region follow a resonant path around the electric field minimum. As we have already discussed, if the gain in kinetic energy ($g$) is such that after the collision with the wall, yield of secondary emission is >1, the multipacting will occur. If the elliptical cavity is modestly deformed near the equator region, field pattern nearby changes, and consequently the multipacting growth rates get affected. This slight change in geometry will keep the rf properties of the cavity nearly unchanged. In our case, we deformed the equator region by adding a convex protrusion ( radius ~ 10 mm) along the weld line of cavity [7]. An representational drawing of the geometry modification is shown in Figure 12. We have simulated multiapcting in the geometrically modified cavities with different radii of convex deformations. We observe that, for radius of convexity >4 mm, multipacting does not occur. Though we can select any value of radius of convex deformation >4 mm, its value should be decided by existing manufacturing tolerances.

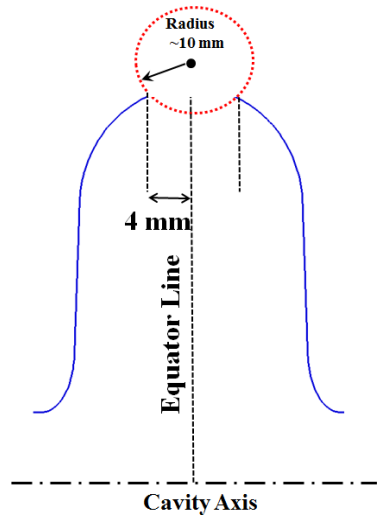

*Figure 12*: A schematic diagram for modification of geometry near the equtor region. A convexity of radius ~10 mm is introduced near the equator region.  This convexity is spread in the width of ~8 mm around the equator line.

Growth rate of multipacting electrons ($dn/dt$) is calculated as a function of $E_{acc}$ for geometrically modified cavity for $\beta_g$ =0.61 elliptical cavity, as shown in Figure 11. We can see that multipacting growth rate becomes negative and therefore multipacting does not exist for the modified geometry. Similarly, for $\beta_g$ =0.9 cavity, the effect of deformation (convexity with radius ~ 10 mm) near the equator region of geometry can be seen in Figure 13. Here, multipacting is again completely non-existent for the modified cavity. Multipacting analysis for modified geometry (~ 10 mm convex deformation) of $\beta_g$ =0.61 and $\beta_g$ =0.9 cavities was also done using `CST-PIC`. Obtained results show that multipacting ranges of $E_{acc}$ are reduced in comparison to unmodified geometry.

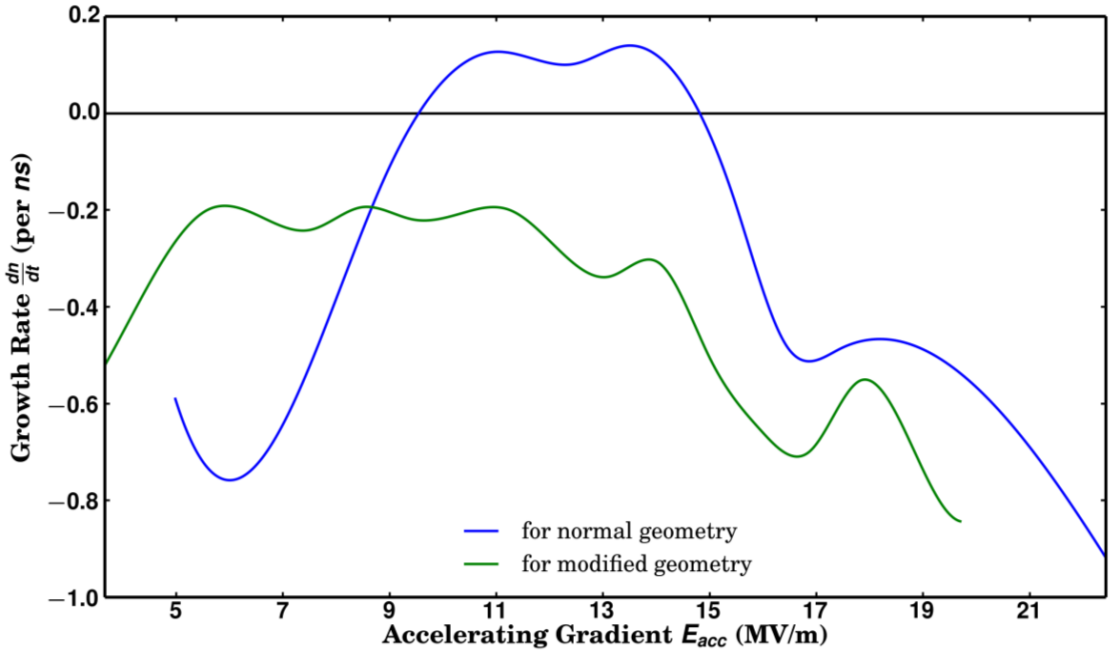

*Figure 13*: Growth rate ($dn/dt$) of multipacting electrons is plotted as a function of $E_{acc}$. The curve in blue color shows the growth rate for unmodified geometry of the $\beta_g$ =0.9 elliptical cavity. The curve in green color represents the growth rate when the cavity geometry modified by small convex perturbation near the equator region. These results are obtained using the indigenously developed code.

## 6. Discussions and Conclusions

In this paper, we have analyzed the phenomenon of multipacting in niobium based SRF cavities in a detailed manner. In order to perform the multipacting analysis, we have developed a computer code, which was used to analyze the multipacting phenomenon in $\beta_g = 0.61$ and $\beta_g = 0.9$, 650 MHz elliptical cavities that will be used in the medium and high energy section of 1 GeV H$^-$ injector linac for the proposed ISNS project.

The stepwise procedure that we have used in our code for the analysis of multipacting is summarised as follows-

1. We start the multipacting analysis by determining the multipacting prone location (equator region) in the elliptical cavity. In order to determine the multipacting prone location, we track a randomly initialized particle for few rf cycles and observe that it slowly drifts towards the equator region.

2. In this step, we search for the range of $E_{acc}$ for which resonant trajectories are possible using a tracking subroutine that we have developed. To determine the range of $E_{acc}$, we vary the $E_{acc}$ and the initial phase $\phi_0$, and track the particle trajectories up to ~20 successive collisions with the cavity wall. The range of $E_{acc}$ is then determined for which the average phase difference between two successive collisions is $\pi$. In this range of $E_{acc}$, the particle follows resonance condition for two point first order multipacting. Since two point first order multipacting is prominent in the TESLA type cavities, we focus our computational effort in this range of $E_{acc}$.

3. In the next step, along with the tracking subroutine, we run another subroutine, which we have developed based on Furman Model, to generate the secondary emission . These two subroutines run sequentially in a repeated manner to ultimately yield the number of multipacting electrons. We then plot the growth rate of multipacting electrons as a function of $E_{acc}$ and identify the range of $E_{acc}$ for which the growth rate is positive.

Next, we benchmarked our code with the experimental results published in Ref. [7]. Then we perform the multipacting analysis for $\beta_g$ =0.61 and $\beta_g$ =0.9, 650 MHz elliptical cavities in order to estimate the range of $E_{acc}$ for which the multipacting will occur. Also, we have shown that if we refine the cavity geometry by introducing a small convex curvature of ~ 10 mm near the equator region, multipacting will be reduced significantly. In comparison to the results obtained from `CST-PIC` simulation, we have observed that results obtained using our code agree more closely with the experimental results. Though a range of 2D and 3D codes are available for multipacting studies, most of the 2D codes follow a simplistic secondary emission model and are unable to detect multipacting in elliptical geometry. On the other hand, contemporary 3D codes like `CST` are able to detect multipacting, but they are extremely time and resource consuming. In comparison to this scenario, our code which is a 2D code, but incorporates the sophisticated Furman Model and can detect multipacting in the elliptical cavity. Being a 2D code, the code has to handle less data and is therefore fast. In other words, we can say that our code is uniquely equipped with speed of 2D codes and accuracy of results of 3D codes (*e.g.* `CST`). As the elliptical SRF cavities are axially symmetric, our code is suitable in analyzing the multipacting phenomena in the most efficient way. However, the code can be extended to 3D as well.

In Ref. [7], it has been observed that computer code `Multipac` is unable to detect multipacting in the elliptical cavities while experimental observations show that multipacting occurs in these cavities. It is shown in this reference that multipacting can be predicted using CST-PIC in the same geometry. The reason for this contradictory result as described in Ref [7] is that `Multipac` uses a simple secondary emission model where emission energy of the secondary electron is taken as a constant value. Also Ref. [7] points out that the other drawback of the emission model used in `Multipac` is that it ignores the effect of elastic and rediffused type of secondary emissions from the surface. The multipacting can be predicted by the computer codes like CST-PIC which use the secondary emission model incorporating these emission processes. To reinforce this observation, in Ref. [7], it has been shown that CST-PIC cannot detect multipacting if elastic and rediffused emissions are turned off. However, our observation with Nb surface shows that elastic and rediffused process contribute a very small fraction of total secondary emission. The secondary emission in this case mainly occurs through the true-secondary process, as shown in Figure 14. Since elastic and rediffused secondary emission have very small contributions (<1% for $E_0$ > 10 eV), these factors cannot be the reason for the non-detection of multipacting on the Nb surface.

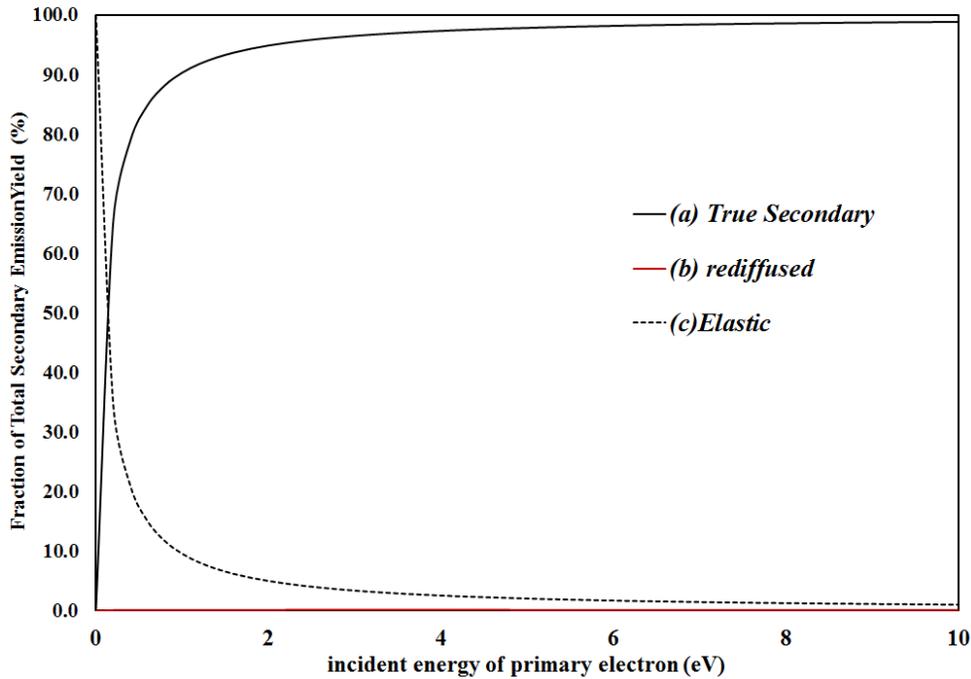

*Figure 14*: Contribution to the total secondary emission yield from the three components- (a) true secondary emission, (b) rediffused emission and (c) elastic emission is plotted as a function of collision energy of primary electrons at normal incidence.

Using indigenously developed computer code, we carried out the multipacting analysis of the $\beta_g$ =0.61, 650 MHz elliptical cavity. We have calculated the multipacting growth rate using Furman model for two cases – in the first case, contributions in secondary emission were taken from all three processes, in the second case, we turned off the contributions from the elastic and rediffused processes. In Figure 15, we can see that results obatined in both the cases are similar. This shows that the main reason for detection of multipacting on the Nb surface is the presence of true secondary emission process in the secondary emission model.

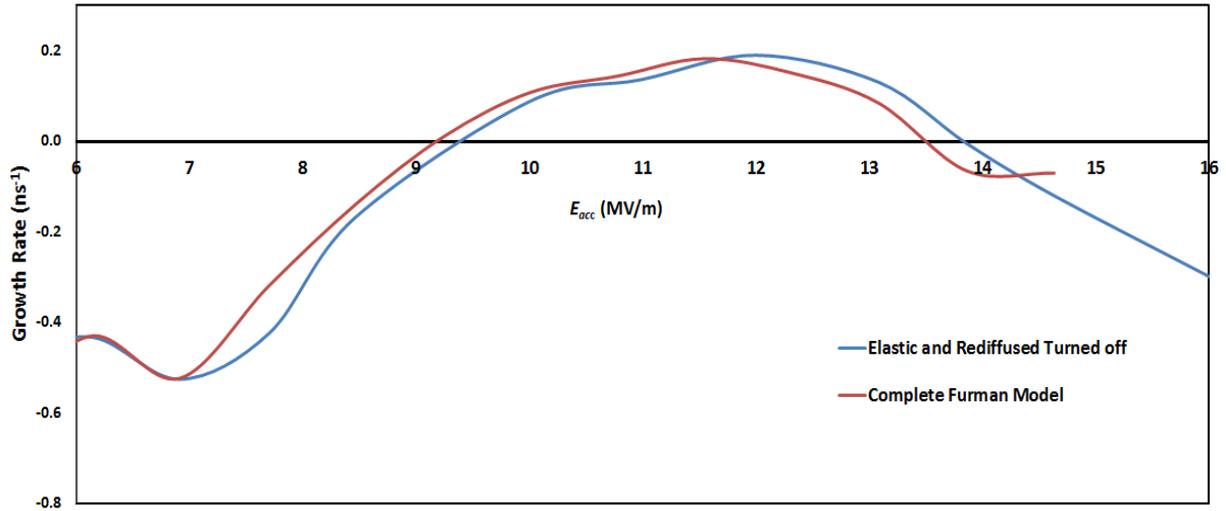

*Figure 15* : Growth rate of multipacting electrons is plotted as a function of $E_{acc}$ for $\beta_g$ =0.61, 650 MHz elliptical cavity. These results are obtained for the two cases- (a) when all three processes contribute to the secondary emission in indigenously developed code, and (b) when elastic and rediffused contribution are turned off in the code.

To conclude, in this paper, we have described a procedure to analyze the multipacting in a step by step method. Also we present the complete multipacting analysis following these steps using an efficient code which we have developed specially for axially symmetric cavities. Finally, we presented the multipacting analysis of $\beta_g$ =0.61 and $\beta_g$ =0.9, 650 MHz elliptical cavity which will be used in the medium and high energy section of the proposed ISNS linac. The analysis presented in this paper and the computer code developed by us will be useful in designing SRF cavities for various applications.

# References

[1] J. Knobloch, W. Hartung, and H. Padamsee. "Multipacting in 1.5-GHz superconducting niobium cavities of the CEBAF shape" Presented at the *8th Workshop on RF Superconductivity*, Padova, Italy, SRF 981012-09, 1997

[2] C. M. Lyneis, H. A. Schwettman and J. P. Turneaure, "Elimination of electron multipacting in superconducting structures for electron accelerators ", Applied Physics Letters, 31, 541 (1977).

[3] B. Aune et al. "Superconducting TESLA cavities" *Physical Review Special Topics - Accelerators And Beams*, Volume 3, 092001 (2000)

[4] H. Padamsee, J. Knobloch, and T. Hays, RF superconductivity for Accelerators (Wiley publishing, 1998).

[5] PASI YLA-OIJALA, Electron Multipacting In Tesla Cavities And Input Couplers, *Particle Accelerators*, Vol. 63, pp. 105-137, 1999

[6] Valery Shemlin. "Multipactor in crossed rf fields on the cavity equator " *Physical Review Special Topics - Accelerators And Beams*, Volume 16, 012002 (2013)

[7] S. Kazakov, I. Gonin, and V. Yakovlev, in *Proceedings of IPAC2012, New Orleans, Louisiana, USA* (2012).

[8] F.L.Krawczyk, in *The 10th Workshop on RF Superconductivity, 2001, Tsukuba, Japan* (Los Alamos National Laboratory, Los Alamos, NM, USA, 2001).

[9] P. Yla-Oijala and D. Proch, in *The 10th Workshop on RF Superconductivity, 2001, Tsukuba, Japan* (2001).

[10] WWW.CST.COM.

[11] M. A. Furman, Tech. Rep., centre for beam physics, acclerator and fusion research division, Lawrence Berkley National laboratory (2002).

[12] H. Bruining, "*Physics and applications of Secondary Electron Emission*" (McGraw HILL Book Co., New York, 1956)

[13] www.python.org

[14] D. Griffith, *Introduction to Electrodynamics* (Prentice Hall, 2009).

[15] R.S. Palais, R.A. Palais, Differential Equations, Mechanic, and Computation, American Mathematical Society, USA, 2009.

[16] C. K. Birdsal and A. B. langdon, *Plasma Physics via Computer Simulation* (The adam Hilger series on Plasma Physics, 1991).

[17] V. S. S. Belomestnykh, Nuclear Instruments and Methods in Physics Research A **595**, 293 (2008).

[18] R. A. Kishek, Physical Review Letters (2012).

[19] S. C. Chapra, *Numerical Methods For Engineers* (McGraw Hill Publication, 2006).

[20] A. R. Jana and V. Kumar, IEEE Transactions on Applied Superconductivity (2013a).

[21] A. R. Jana and V. Kumar, IEEE Transactions on Applied Superconductivity (2013b).